\documentclass[12pt]{article}
\usepackage{amsmath}        % not available on your system
\usepackage{amsthm}
\usepackage{subfigure}
\usepackage{amsfonts}
\usepackage{makeidx}         % allows index generation
\usepackage{graphicx}        % standard LaTeX graphics tool
\usepackage{multicol}        % used for the two-column index
\usepackage[bottom]{footmisc}% places footnotes at page bottom
\usepackage{color}
\newtheorem{lemma}{Lemma}

\newtheorem{rem}{Remark}

\begin{document}

\begin{center}{\large\sc A goodness of fit test for the Pareto distribution}
{   %
\\ 
} \vspace*{0.5cm} {Emanuele Taufer$^{{\rm{a}}}$, Flavio Santi$^{{\rm{a}}}$, Giuseppe Espa$^{{\rm{a}}}$, Maria Michela Dickson$^{{\rm{b}}}$,  }\\ \vspace*{0.5cm}
{\it 
$^{{\rm{a}}}$Department of Economics and Management, University of Trento - Italy 
}
{\it 
$^{{\rm{a}}}$Department of Statistics, University of Padua - Italy 
}
\vspace*{0.5cm} 

%\today
\end{center}
\abstract{The Zenga (1984) inequality curve $\lambda(p)$  is constant in $p$ for Type I Pareto distributions. This characterizing behavior will be exploited to obtain graphical and analytical tools for tail analysis and goodness of fit tests. A testing procedure for Pareto-type behavior  based on a regression of $\lambda(p)$  against $p$ will be introduced. 
}
\medskip


\baselineskip = 1.1\baselineskip

\section{Introduction}
\label{sec:1}

Let $X$ be a positive random variable with finite mean $\mu$, distribution function $F$, and probability density $f$. The inequality curve, $\lambda(p)$ , defined in \cite{Zenga84}  is defined as:
\begin{equation}\label{lambdap}
\lambda(p) = 1 - \frac{\log (1-Q(F^{-1}(p)))}{\log(1-p)}, \quad 0<p<1,
\end{equation}
where $F^{-1}(p)= \inf \{x\colon F(x) \geq p \}$ is the generalized inverse of $F$ and $Q(x) = \int_0^x t f(t) dt / \mu$ is the first incomplete moment. $Q$ can be defined as a function of $p$ via the Lorenz curve
\begin{equation}
L(p)=Q(F^{-1}(p)) =\frac{1}{\mu} \int_0^p F^{-1}(t) dt.
\end{equation}

$\lambda(p)$  can be used to define a concentration measure as it has been done in \cite{Zenga84}. Here we exploit the  curve in order to define goodness-of-fit test for the Pareto. In fact as it will be more formally shown below $\lambda(p)$ is constant in $p$ for type I Pareto distributions. Indeed the above properties can also be exploited in order to define graphical tools for the analysis of distributions and their tails.  For related works see  \cite{Mei15}, \cite{Vol16}, \cite{Grahovacetal2015}, \cite{Jia2018}, \cite{Leo06},\cite{Leo13}, \cite{mcneil2005}, \cite{resnick1999}. 

For a Type I Pareto  distribution \cite[573~ff.]{johnson1995} with
\begin{equation}\label{ParetoF}
F(x)=1-(x/x_0)^{-\alpha}, \quad x \geq x_0
\end{equation}
it holds that $\lambda(p)=1/\alpha$, i.e.\ $\lambda(p)$ is constant in $p$. This is actually an if-and-only-if result, as we formalize in the following lemma:

\begin{lemma}
\label{lem:iff}
The curve $\lambda(p)$ defined in \eqref{lambdap} is constant in $p$ if, and only if, $F$ satisfies~\eqref{ParetoF}.
\end{lemma}

\section{Goodness-of-fit tests}

Let $X_{(1)}, \dots, X_{(n)}$ be the order statistics of the sample, $\mathbb{I}_{(A)}$ the indicator function of the event $A$. To estimate $\lambda(p)$, define the preliminary estimates
\begin{equation}
F_n(x)= \frac{1}{n} \sum_{i=1}^n \mathbb{I}_{(X_i \leq x)}  \qquad Q_n(x) = \frac{\sum_{i=1}^n X_i \mathbb{I}_{(X_i \leq x)}}{\sum_{i=1}^n X_{i}}
\end{equation}
Under the Glivenko-Cantelli theorem (see e.g.~\cite{resnick1999}) it holds that $F_n(x) \to F(x)$ almost surely and uniformly in $0<x<\infty$; under the assumption that $E(X) < \infty$, it holds that $Q_n(x) \to Q(x)$ almost surely and uniformly in $0<x<\infty$.  $F_n$ and $Q_n$ are both step functions with jumps at $X_{(1)}, \dots, X_{(n)}$. The jumps of $F_n$ are of size $1/n$ while the jumps of $Q_n$ are of size $X_{(i)}/T$ where $T=\sum_{i=1}^n X_{(i)}$. Define the empirical counterpart of $L$ as follows:
\begin{equation}\label{Lnp}
L_n(p) = Q_n( F_n^{-1}(p)) = \frac{\sum_{j=1}^i X_{(j)}}{T}, \quad \frac{i}{n} \leq p < \frac{i+1}{n}, \quad i=1, 2, \dots, n-1, 
\end{equation}
where $F_n^{-1}(p)= \inf \{x: F_n (x) \geq p \}$.  To estimate $\lambda(p)$ define

\begin{equation}
{\hat \lambda}_i = 1- \frac{\log(1-L_n(p_i))}{\log(1-p_i)}, \quad p_i = \frac{i}{n}, \quad i= 1, 2, \dots n-\left\lfloor \sqrt{n} \right\rfloor.
\end{equation}
The choice of $i = 1, \dots , n - \left\lfloor \sqrt{n} \right\rfloor$ guarantees that  $\hat\lambda_i$ is consistent for $\lambda_i$ for each $p_i=i/n$ as $n \to \infty$.

Goodness-of-fit tests can be defined by linear regression of $\lambda_i$, on $p_i$.  From Lemma \ref{lem:iff},  for a distribution $F$ satisfying  ~\eqref{ParetoF} with $\alpha>1$, for any choice of $p_i$, $0<p_i<1$, $i=1, \dots, m$,  one has the linear equation
\begin{equation}
\lambda_i= \beta_0+ \beta_1 p_i,
\end{equation}
where $\beta_0=1/\alpha$ and $\beta_1=0$.  Given a random sample $X_1, \dots , X_n$, estimation and testing procedures can be defined through the regression 
\begin{equation}
\hat \lambda_i = \beta_0 + \beta_1 p_i + \varepsilon_i
\end{equation} 
where $\varepsilon_i=\hat{\lambda}_i-\lambda_i$. Hence an estimator of $\beta_0$ can be used to estimate $\alpha$ while a test on the hypothesis $H_0: \beta_1=0$ can be used to test that a distribution $F$ satisfies ~\eqref{ParetoF}.

Using least squares estimators and exploiting the knowledge that $\beta_1=0$ in the estimation of $\beta_0$, define
\begin{equation}
 \hat \beta_0 = \frac{1}{m}\sum_{i=1}^m \hat \lambda_i, \qquad \hat \beta_1 = \sum_{i=1}^m \frac{\hat \lambda_i (p_i - \bar{p})}{S^2_p}=\sum_i^m \hat \lambda_i c(p_i)
\end{equation}
where $\bar{p}$ is the mean of the $p_i$'s and $S^2_p=\sum_i^m(p_i-\bar{p})^2$, $c(p_i)= (p_i - \bar{p})/S^2_p$.

Note that since
$$\sum_{i=1}^n i = \frac{n(n+1)}{2} \quad \text{and}\quad \sum_{i=1}^n i^2 = \frac{n(n+1)(2n+1)}{6}$$
then, for $p_i = i/n$, $i= 1, \dots, m$,  $m=n - \left\lfloor \sqrt{n} \right\rfloor$,
$$
\bar{p} = \frac{1}{2} \frac{m(m+1)}{n^2} = \frac{1}{2}+O\left( \frac{1}{\sqrt{n}}\right) \quad \text{and} \quad S^2_p = \frac{1}{12} \frac{m(m^2-1)}{n^2}=O(n).
$$

\begin{rem}
Since $\lambda(p)$  does not depend on location parameters, one can construct goodness-of-fit tests free of $x_0$ for the Pareto distribution.
\end{rem}

A formal test for the general hypothesis that the data comes from a distribution $F$ satisfying \ref{ParetoF}, i.e.
$$
H_0: F \quad \text{is} \quad Pa(\alpha, x_0), \quad \alpha> 1, \quad x_0>0
$$ 
The null hypotheses is rejected if $|\hat \beta_1|$ is large. In order to carry on practically the test we have two possibilities. The first is to use a normal approximation to $\hat \beta_1$, properly normalized. This is feasible only for the cases $\alpha >2$. 

The second way is to carry on a parametric bootstrap procedure as follows:

\begin{itemize}

\item[1.] Given a random sample of size $n$, estimate  $\hat \alpha=1/\hat \beta_0$ and  $\hat \beta_1$.

\item[2.] Generate  a sample of size $n$ from a Pa$(\hat \alpha,1)$and estimate $\hat \beta_1$. Note that since $\lambda(p)$ does not depend on $x_0$, we do not need to estimate it and use, for example, always the same value $1$.

\item[3.] Repeat step 2 $M$ times.

\item[4.] Get an estimated $p$-value of $\hat \beta_1$ from the bootstrap distribution.
\end{itemize} 

To compare the performance of the test proposed here, consider \cite{Vol16}. For the distributions and sample sizes considered in Table 8 of \cite{Vol16}, Table \ref{tab:GoF1} contains the power estimates obtained with the parametric bootstrap for tests of level 0.05. Results are based on 500 samples of size $n$ from null and alternative distributions; for each of them a parametric bootstrap with $M=500$ was carried on.  

\begin{table}[ht]

 \centering
 {\small
 \begin{tabular}{rrrrrrrrr}
   \hline
 n & $Pa(2)$ & $LN(1)$ & $LN(2.5)$ & $LN(3)$ & $Exp$ & $Ga(2)$ & $LW(0.25)$ & $LW(0.5)$\\ 
20 &         &0.19&      &      &0.99&0.99&0.23&0.46 \\
50 &0.054&1.00&0.19&0.04&1.00&1.00&0.38&0.81 \\
100&0.048&1.00&0.61&0.03&1.00&1.00&0.59&0.96 \\
500&0.038&1.00&0.81&0.08&1.00&1.00&0.96&1.00\\
1000&0.044&1.00&0.95&0.12&1.00&1.00&0.98&1.00 \\

   \hline

    \hline
 \end{tabular}
 }
\caption{\footnotesize Estimated power for tests of size 0.05. Results based on 500 replications, $p$-value estimates were obtained by parametric bootstrap with $M=500$.}
	\label{tab:GoF1}

 \end{table}

We see that the test proposed here performs better than its competitors in several cases. The Log-normal distribution looks like a hard alternative for large values of the standard deviation. Further simulation and analyses will be carried on in a subsequent work.

\newpage
%%%%%%%%%%%%%%%%%%%%%%%% referenc.tex %%%%%%%%%%%%%%%%%%%%%%%%%%%%%%
% sample references
% %
% Use this file as a template for your own input.
%
%%%%%%%%%%%%%%%%%%%%%%%% Springer-Verlag %%%%%%%%%%%%%%%%%%%%%%%%%%
%
% BibTeX users please use
% \bibliographystyle{}
% \bibliography{}
%

\end{document}